\documentstyle[12pt]{article}
\def\journal#1, #2, #3, #4 { {\sl #1~}{\bf #2~} (#3)  #4 }

\def\prd{\journal Phys. Rev. D, }

\def\prl{\journal Phys. Rev. Lett., }

\def\cmp{\journal Comm. Math. Phys., }

\def\np{\journal Nucl. Phys., }

\def\pl{\journal Phys. Lett., }

\def\annp{\journal Ann. Phys. (N.Y.), }
\def\ijmp{\journal Int. J. Mod. Phys., }

\catcode`\@=11
\def\marginnote#1{}
\newcount\hour
\newcount\minute
\newtoks\amorpm
\hour=\time\divide\hour by60
\minute=\time{\multiply\hour by60 \global\advance\minute
by-\hour}\edef\standardtime{{\ifnum\hour<12
\global\amorpm={am}%
        \else\global\amorpm={pm}\advance\hour by-12 \fi
        \ifnum\hour=0 \hour=12 \fi
        \number\hour:\ifnum\minute<10
0\fi\number\minute\the\amorpm}}
\edef\militarytime{\number\hour:\ifnum\minute<10
0\fi\number\minute}

\def\draftlabel#1{{\@bsphack\if@filesw {\let\thepage\relax
   \xdef\@gtempa{\write\@auxout{\string
      \newlabel{#1}{{\@currentlabel}{\thepage}}}}}\@gtempa
   \if@nobreak \ifvmode\nobreak\fi\fi\fi\@esphack}
        \gdef\@eqnlabel{#1}}
\def\@eqnlabel{}
\def\@vacuum{}
\def\draftmarginnote#1{\marginpar{\raggedright\scriptsize\tt#1}}
\def\draft{\oddsidemargin -.5truein
        \def\@oddfoot{\sl preliminary draft \hfil
        \rm\thepage\hfil\sl\today\quad\militarytime}
        \let\@evenfoot\@oddfoot \overfullrule 3pt
        \let\label=\draftlabel
        \let\marginnote=\draftmarginnote

\def\@eqnnum{(\theequation)\rlap{\kern\marginparsep\tt\@eqnlabel}%
\global\let\@eqnlabel\@vacuum}  }

\def\numberbysection{\@addtoreset{equation}{section}
        \def\theequation{\thesection.\arabic{equation}}}

\def\underline#1{\relax\ifmmode\@@underline#1\else
 $\@@underline{\hbox{#1}}$\relax\fi}

\catcode`@=12
\relax

\numberbysection

\def\beq{\begin{equation}}
\def\eeq{\end{equation}}
\def\beqa{\begin{eqnarray}}
\def\eeqa{\end{eqnarray}}
 \def\nnn{\nonumber \\}

\def\Jhat{{\widehat J}}

\def\mhat{{\widehat m}}

\def\xib{{\overline \xi}}

\def\qhat{{\widehat q}}

\def\mhat{{\widehat m}}

\def\Mb{{\bar M}}
\def\sb{{\bar s}}
\def\zb{{\bar z}}

\def\etab{\bar \eta}
\def\Mc{{\cal M}}
\def\Mcb{{\bar {\cal M}}}
\begin{document}
\tolerance 2000
\hbadness 2000
\begin{titlepage}

\nopagebreak \begin{flushright}

LPTENS--93/30\\
hep-th@xxx/9308134
 \\
    August  1993
\end{flushright}

\begin{center}
{\large \bf
THE MANY FACES OF THE   \\
\medskip
QUANTUM LIOUVILLE EXPONENTIALS } \\
\bigskip
{\bf Jean-Loup~GERVAIS}\\
\medskip
and \\
\medskip
{\bf Jens SCHNITTGER}{\footnotesize\footnote{supported by DFG}}\
\\medskip
{\footnotesize Laboratoire de Physique Th\'eorique de
l'\'Ecole Normale Sup\'erieure\footnote{Unit\'e Propre du
Centre National de la Recherche Scientifique,
associ\'ee \`a l'\'Ecole Normale Sup\'erieure et \`a
l'Universit\'e
de Paris-Sud.},\\
24 rue Lhomond, 75231 Paris CEDEX 05, ~France}.
\end{center}
\vglue 1 true cm
\begin{abstract}
\baselineskip .4 true cm
{\footnotesize
\noindent
  First,
it is proven that the three main operator-approaches to the
quantum Liouville exponentials
 --- that is the one of Gervais-Neveu
(more recently developed further by Gervais),
Braaten-Curtright-Ghandour-Thorn,
and Otto-Weigt --- are equivalent since they are related by simple
basis transformations in the Fock space of the free field  depending
upon the zero-mode only. Second,
the GN-G expressions for quantum Liouville
exponentials, where the
$U_q(sl(2))$ quantum group structure is
manifest, are  shown to be  given
by  q-binomial sums over powers of the chiral fields in the $J=1/2$
representation. Third, the Liouville exponentials are expressed as operator
tau functions whose chiral expansion exhibits a q Gauss decomposition,
which is the direct quantum group analogue of the classical solution of
Leznov and  Saveliev. It involves q exponentials of quantum group generators
with group ``parameters'' equal to chiral components of the quantum metric.
Fourth, we point out that the OPE of the $J=1/2$ Liouville exponential
provides the quantum version of the Hirota bilinear equation.
}
\end{abstract}
\vfill
\end{titlepage}
\section {INTRODUCTION}
The problem of understanding the structure of 2d gravity from
the continuum
point of view can look back upon a history of more than ten years.
During this time,
various frameworks have been put forward to construct
the quantum Liouville field,
either in a canonical setup, or more recently, by path integral	 methods.
For the classical theory itself, our understanding
was much deepened by studying its generalization to
  the  infinite (Toda)  hierarchy of integrable systems  where the
powerful machinery of the Kyoto school is at work.

The main\footnote{
We consider only the case of finite volume ---
which is relevant e.g. for the
string application --- and thus will not  discuss  the approaches
of refs.\cite{MF,HJ}.} quantum canonical schemes\cite{GN,BCGT,OW}
 that have been  proposed\footnote{The recent article of
Kazama and Nicolai\cite{KN}  follows  the
scheme of ref.\cite{OW}, so that we do not treat it separately.}
look rather different
at first glance. In particular two of them\cite{BCGT,OW} aimed at
establishing quantum B\"acklund-type transformations into  a particular
free field, while
the third\cite{GN}  worked more symmetrically, and
 developed  an operatorial
scheme that is close to the spirit of the BPZ method.
It was more recently realized\cite{B,G}  that the  basic
principle behind this third
scheme (ref.\cite{GN})    is
the  $U_q(sl(2))$ quantum-group structure of the theory, which determines
the form of the relevant chiral braiding and fusing matrices in terms
of quantum group symbols. This quantum group aspect
seems  rather remote at first sight from the  free-field calculations
of the other two (refs.\cite{BCGT,OW})  and
the question   arises,  whether,  and in
which  sense the  three schemes just mentioned
are equivalent.
This problem is also of relevance
for the relation
between the continuum and matrix model approaches, in view of the
recent result\cite{G5} that
the approach of refs.\cite{GN,G} is    able to
reproduce the  matrix-model three-point functions.
 In the first part of this article, we
shall show that  the powers
of the two-dimensional metric do  indeed
agree in all three approaches up to
a simple equivalence transformation.
In the second part we push further the quantum group aspect of the
approach of refs.\cite{GN,G}. Indeed, simple manipulations of the
expression of the Liouville exponential derived in ref.\cite{G} in terms
of quantum group covariant chiral vertex operators will allow us to
rewrite it as a q binomial expansion, or
 as a kind  of operator tau-function that obeys
 an operator Hirota equation,
in close parallel to the
classical Leznov-Saveliev\cite{LS} solution, and Kyoto school
approach\cite{JM,KW}.

The paper is organized as follows.
Sections 2 and 3   display  a detailed comparison of the three
schemes
of refs.\cite{GN,G,BCGT,OW}.
Since the main ingredients in all constructions are locality
and conformal invariance, the analysis of their equivalence also
provides evidence for the problem
whether these two requirements suffice to determine the structure
of the quantum theory
uniquely. It may seem surprising
that no such investigation has been carried
out to date. In the present analysis we shall demonstrate therefore
in a very explicit manner the equivalence of the approaches put
forward by
Gervais and Neveu\cite{GN} and more recently \footnote{GN
only considered open strings in detail. Therefore we
compare with the more recent expression of ref.\cite{G}.}
by Gervais\cite{G} (GN-G),
Braaten, Curtright, Ghandour  and Thorn\cite{BCGT} (BCGT) and Otto and
Weigt\cite{OW}  (OW).
The relevant conformal objects being exponentials
of the Liouville field, we shall show that there exist what one may
regard as
basis-transformation operators
$S_{1,2}(\varpi )$ depending only on the free-field
zero mode $\varpi $ such that (up to trivial field redefinitions, see
below)
\beq
e^{-J\displaystyle  \varphi}_{GN-G} =
S_1(\varpi )e^{-J\displaystyle  \varphi}_{BCGT}S_1^{-1}
(\varpi )=
S_2(\varpi )e^{-J\displaystyle  \varphi}_{OW}S_2^{-1}(\varpi )
\label{1.1}
\eeq
A correspondence of this type   could be
expected since all three frameworks
are based on a transformation of the Liouville
field $\varphi$ onto a set of
free fields, such that the energy-momentum tensor
takes the same form in all approaches, and impose the same
locality condidtion.
 However, the unexpected point is that
{\bf the operators $S_\ell$ only involve the zero-mode}. Thus the
correspondence is basically rather simple, and
as a consequence,   the exponentials
 can all be written in the form (see below for details)
\beq
e^{-J\displaystyle \varphi} =\sum_{J+m=0,1,2,...} a_m^{(J)}(\varpi) V_m^{(J)}
\bar V_m^{(J)}
\label{1.2}
\eeq
where $a_m^{(J)}(\varpi )$ are suitable coefficients, and
$V_m^{(J)}$ resp. $\bar V_m^{(J)}$ are the same left-  resp. right-moving
primaries in all approaches,
 given in terms of free fields, with a conformal dimensions
characterized by $J$ (equal, of
course, to the dimension of the exponential itself) and a definite shift
of the zero mode given by $m$:
\beq
\varpi V_m^{(J)}  =V_m^{(J)}\>  (\varpi -2m)
\label{1.3}
\eeq
and likewise for $\bar V_m^{(J)}$.
The equivalence Eq.\ref{1.1} is easily seen by use of Eq.\ref{1.3}
to imply (for integer $2J$)
the following relation between the coefficients $a_m^{(J)}$ in the
3 frameworks:
$$
a_m^{(J)}(\varpi )|_{GN-G} =C_1^{2J}\,\prod_{r=0}^{2m-1} f_1(\varpi +r)
\, a_m^{(J)}(\varpi )|_{BCGT} \,
$$
\beq
= C_2^{2J}\,\prod_{r=0}^{2m-1} f_2(\varpi +r)
\, a_m^{(J)}(\varpi )|_{OW}
\label{1.4}
\eeq
with suitable functions $f_{1,2}(\varpi )$. In Eq.\ref{1.4}, we have
accounted also for the possibility of trivial redefinitions $\varphi
\rightarrow \varphi +\ln C^2$; this is the origin of the
constants $C_1,C_2$. The products are defined
in the standard fashion when $2m$ is a negative integer
(see e.g. ref.\cite{G}). Using the recently obtained
result\cite{GSch} that the GN-G construction can be extended to
noninteger $2J$, we shall show that the OW and GN-G operators are
actually equivalent for arbitrary $J$ (in the BCGT framework, only
the exponential with $J=1/2$ was constructed explicitly).
The correspondences  BCGT vs. GN-G, and
OW vs. GN-G  are  the subject of
sections  2, and 3, respectively.  To facilitate
comparison with the literature, the notations of BCGT resp. OW will
be used in the corresponding sections, with
comprehensive translation tables
presented in  an appendix. In the rest of the paper we use the notation
of GN-G.


In section 4, we next
briefly  recall the group theoretical approach to the classical
Liouville exponential, based on the solution of ref.\cite{LS}.
The exponential $e^{-\varphi /2}$ describing
the inverse square root of the
metric plays a special role for the integrability structure of the theory.
In the standard Lax pair approach, it appears as the monodromy invariant
solution of the auxiliary linear system\cite{B,GN6}. We use the
group-theoretic viewpoint to derive a bilinear equation
of the Hirota type for this
exponential  --- of course equivalent  to the
Liouville equation ---   which clearly  shows  that
it is a  tau function.
In section 5, we start from an expression of  the Liouville
exponential in terms of a different set of fields $\xi_M^{(J)}$
that are quantum group
covariant.  First we show that the expansion of the
quantum Liouville exponential in terms of the $\xi$
fields is a field-theory analogue of  the expansion of a
q binomial.  Then a group-theoretic method is developed, based on the
q deformation of the classical $sl(2)$  group of section 3, which is shown
to be directly connected with the $U_q(sl(2)$ quantum group structure
of the GN-G approach. The quantum Liouville exponentials are written
using a generalized Gauss decomposition of the quantum group
where the group ``parameters'' are  the covariant $\xi$ fields
just mentioned. This shows how
 the quantum Liouville exponential may be regarded as a q tau
function --- which is an operator.
The quantum Hirota equation is established by  observing
that the quantum equivalent of the classical bilinear equations is  the
fact that  a particular term in  the short-distance operator-product
expansion of the quantum inverse square root   of the
metric is a  constant, since
it  is  given by its  zeroth power. From this
viewpoint
the  derivation of the quantum Hirota equation
 becomes  a straightforward consequence of the relation
between Clebsch-Gordan coefficients and fusing matrix of the
$\xi$ fields.

\section{ BCGT vs. GN-G}

In the BCGT formalism, the Liouville field is expressed in terms of free
chiral fields $\psi_L ,\psi_R $ by means of a B\"acklund transformation.
On the quantum level, the authors were able to explicitly
construct
the inverse square root of the metric, but could only
obtain  approximate
results for other Liouville exponentials. Following ref.\cite{BCGT},
this particular power  is represented as
\beq
e^{\displaystyle -g\Phi (\tau ,\sigma) } =\zeta m e^{-g\tilde \psi^-
(\tau ,\sigma)}
Z(\tau, \sigma) e^{-g\tilde \psi^+ (\tau ,\sigma )}
\label{2.1}
\eeq
Here,
the fields $\tilde \psi^\pm$ are just the  annihilation
resp. creation parts of
\beq
\tilde \psi (\tau ,\sigma) =\psi_L(\tau +\sigma )-\psi_R(\sigma -\tau)
\label{2.2}
\eeq
with \hfill
$$
\psi_L (\tau +\sigma )={i\over \sqrt {4\pi }}\sum_{n\ne 0} {A_n \over n}e^{
-in(\tau +\sigma )}, \qquad [A_n,A_m ]=n\delta_{n,-m}
$$
\beq
\psi_ R(\tau -\sigma )={i\over \sqrt {4\pi }}\sum_{n\ne 0} {B_n \over n}e^{
-in(\tau -\sigma )}, \qquad [B_n,B_m ]=n\delta_{n,-m}
\label{2.3}
\eeq
The definition of the constants $\zeta$, $m$ and $g$ is given in
 an appendix,
and $\tau ,\sigma $ are world sheet coordinates on the cylinder as usual.
The nontrivial part of the structure of $e^{-g\Phi } $ resides in the
nonlocal periodic operator $Z(\tau ,\sigma)$. It
can be written as an integral over free field exponentials
of dimension 1:
$$
Z(\tau ,\sigma)=\int_0^{2\pi}d\sigma '  f(\sigma -\sigma ')
:F(P)e^{gPd(\sigma '-\sigma )}\cosh (g\psi(\tau ,\sigma'))e^{g\tilde \psi
(\tau ,\sigma')} :
$$
where \hfill
$$
\psi (\tau ,\sigma ) =Q+P{\tau \over 2\pi }+\psi_L (\tau +\sigma )
+\psi_R (\tau -\sigma ), \qquad [Q,P]=i
$$
$$
F(P)=(\sinh ^2(gP/2) +\sin ^2(g^2/4))^{-1/2}
$$
$$
f(\sigma -\sigma' )=\left (4\sin^2{\sigma -\sigma' \over 2}\right )
^{g^2 / 4\pi }
$$
and $d(\sigma -\sigma ')$ is a periodic antisymmetric function given by \hfill
\beq
d(\sigma'-\sigma )={\sigma'-\sigma \over 2\pi}-{1\over 2}\epsilon (\sigma'
-\sigma ) \ ,
\label{2.4}
\eeq
with $\epsilon (\sigma'-\sigma )$ the stair-step function, equal to
the sign of $\sigma'-\sigma$ for $\sigma'-\sigma \in [-2\pi,2\pi]$.
The normal ordering prescription is just the standard one for the harmonic
oscillator modes, whereas for the zero modes one has
\beq
:e^{\alpha Q}g(P): \equiv  e^{\alpha Q/2} g(P) e^{\alpha Q/2}
\label{2.5}
\eeq
(hermitian normal ordering). Eq.\ref{2.1} is fully normal-ordered. We shall
rewrite it as a product of normal-ordered operators which are primary
fields and periodic up to a multiplicative constant. Since everything
is given in terms of chiral free fields, we can w.l.o.g.
put $\tau =0$ in the following. Decomposing the cosh and reorganizing
the zero modes, we have
$$
e^{{\textstyle -g\Phi (\tau ,\sigma )}}=\zeta m \int_0^{2\pi}d\sigma '
({1\over 2} Y_L(\sigma ,\sigma') +{1\over 2}
Y_R(\sigma ,\sigma'))\, ,
$$
$$
Y_L(\sigma ,\sigma')=f(\sigma -\sigma' )
F(P+ig/2)e^{g(P+ig/2)d(\sigma'-\sigma )}e^{+gQ}
:e^{g\psi_R(\sigma )}e^{-g\psi_L(\sigma)} e^{+2g\psi_L(\sigma')}:
$$
\beq
Y_R(\sigma ,\sigma')=f(\sigma -\sigma' )
F(P-ig/2)e^{g(P-ig/2)d(\sigma'-\sigma )}e^{-gQ}
:e^{g\psi_R(\sigma )}e^{-g\psi_L(\sigma)} e^{-2g\psi_R(\sigma')}:
\label{2.6}
\eeq
Using the relations Eqs.\ref{A.4},\ref{A.6} between the free fields of BCGT and
GN-G,
we see immediately that
$$
:e^{g\psi_R(\sigma )}e^{-g\psi_L(\sigma)} e^{+2g\psi_L(\sigma')}:=
:\bar {\hat V}^{(1/2)}_{1/2}(\sigma) \hat V^{(1/2)}_{-1/2}(\sigma)
 \hat V^{(-1)}_{1}(\sigma'):
$$
\beq
:e^{g\psi_R(\sigma )}e^{-g\psi_L(\sigma)} e^{-2g\psi_R(\sigma')}:=
:\bar {\hat V}^{(1/2)}_{1/2}(\sigma) \hat V^{(1/2)}_{-1/2}(\sigma)
 \bar{ \hat V}^{(-1)}_{-1}(\sigma'):,
\label{2.7}
\eeq
where the hats denote the oscillator parts of the corresponding chiral
primaries. In the following, we discuss only $Y_L(\sigma ,\sigma')$ explicitly;
the treatment of $Y_R$ is totally analogous. The RHS of Eq.\ref{2.7} can be
written as
$$
:\bar {\hat V}^{(1/2)}_{1/2}(\sigma) \hat V^{(1/2)}_{-1/2}(\sigma)
 \hat V^{(-1)}_{1}(\sigma'):=(1-z'/z)^{-g^2/2\pi}
\bar{ \hat V}^{(1/2)}_{1/2}(\sigma) \hat V^{(1/2)}_{-1/2}(\sigma)
 \hat V^{(-1)}_{1}(\sigma')
$$
with\hfill
\beq
z:=e^{i\sigma}\quad
z':=e^{i\sigma'}
\label{2.8}
\eeq
and we have \hfill
\beq
(1-z'/z)^{-g^2/ 2\pi} =\left (4\sin^2{\sigma -\sigma' \over 2}\right )
^{-g^2/4\pi}
e^{-{i\over 2}g^2 d(\sigma'-\sigma )} \ \ .
\label{2.9}
\eeq
The function $f(\sigma -\sigma' )$ is thus absorbed by the removal of
the external normal ordering. It remains to reinstate the zero mode dependence
needed to complete $\bar {\hat V}^{(1/2)}_{1/2}(\sigma), \hat V^{(1/2)}_{-1/2}
(\sigma),\hat V^{(-1)}_{1}(\sigma')$ to the full primary fields.
Here we have to take into account that in the GN-G formalism, one works
in a CFT-adapted notation where
two formally independent sets of zero modes $q_0,p_0$ and $\bar q_0,\bar p_0 $
are used for left and right movers, allowing the chiral factorization
of the basic conformal operators. It is well known that one obtains in this way
the same amplitudes as from the a priori representation with only one set of
zero modes, provided of course one restricts to operators with equal left
and right zero mode shifts and puts $p_0=\bar p_0$ in all matrix elements.
In the Liouville
context, the distinction between $p_0$ and $\bar p_0$ is in fact slightly
more than formal due to the possibility of a winding number in the elliptic
sector \cite{JKM,GN5,LuS}. However, in the hyperbolic sector considered by
BCGT and OW -corresponding to regular solutions of the Liouville equation-
one really has $\bar p_0=p_0$, with real $p_0$. Passing from the BCGT to the
GN-G notation, we thus have to replace in $Y_L$:
\beq
e^{gQ}\rightarrow e^{gQ}e^{g\bar Q} \quad ,\quad e^{gPd(\sigma '-\sigma )}
\rightarrow e^{-gP\epsilon (\sigma' -\sigma )/2}e^{(2\sigma'-\sigma)gP/4\pi}
e^{-\sigma g\bar P /4\pi}
\label{2.10}
\eeq
In view of the relations (cf. Eqs.\ref{A.4},\ref{A.6})
$$
V^{(1/2)}_{-1/2}(\sigma )  =          e^{-gQ-gP\sigma /4\pi }
\hat V^{(1/2)}_{-1/2}(\sigma )
$$
$$
\bar V^{(1/2)}_{+1/2}(\sigma)= e^{+g\bar Q -g\bar P  \sigma /4\pi }
                               \bar {\hat  V}^{(1/2)}_{+1/2}(\sigma)
$$
$$
V^{(-1)}_{1}(\sigma')=e^{+2gQ+gP \sigma' /2\pi }\hat V^{(-1)}_{1}(\sigma')
$$
\beq
\bar V^{(-1)}_{1}(\sigma')=e^{-2g\bar Q +g\bar P  \sigma' /2\pi }
                          \bar { \hat V}^{(-1)}_{1}(\sigma')
\label{2.11}
\eeq
we can then rewrite $Y_L(\sigma ,\sigma' )$ in the form
\beq
Y_L(\sigma ,\sigma')=F(P+ig/2)e^{-{1\over 2}gP \epsilon (\sigma' -\sigma )}
\bar V^{(1/2)}_{1/2}(\sigma) V^{(1/2)}_{-1/2}(\sigma)V^{(-1)}_{1}(\sigma')
\label{2.12}
\eeq
and analogously \hfill
\beq
Y_R(\sigma ,\sigma')=F(\bar P-ig/2)e^{-{1\over 2}g\bar P \epsilon (\sigma'
-\sigma )}
\bar V^{(1/2)}_{1/2}(\sigma) \bar V^{(-1)}_{-1}(\sigma')
V^{(1/2)}_{-1/2}(\sigma )
\label{2.13}
\eeq

Thus we obtain altogether
\beqa
e^{\displaystyle -g\Phi (\sigma )}
={\zeta m \over 2}\int_0^{2\pi} d\sigma'
  &\{ F(P+ig/2)e^{-{1\over 2}gP \epsilon (\sigma'-\sigma )}
\bar V^{(1/2)}_{1/2}(\sigma)V^{(1/2)}_{-1/2}(\sigma)V^{(-1)}_{1}(\sigma')
\nnn
 +&F(\bar P-ig/2)e^{-{1\over 2}g\bar P \epsilon (\sigma'-\sigma )}
\bar V^{(1/2)}_{1/2}(\sigma)
\bar V^{(-1)}_{-1}(\sigma')V^{(1/2)}_{-1/2}(\sigma)  \}
\nnn
\label{2.14}
\eeqa
In view of the periodicity of $e^{-g\Phi }$, we can take  $\sigma \in
[0,2\pi]$. The integrals are then recognized to coincide up to a prefactor
with the screening charges of refs.\cite{LuS,GSch}:
$$
\int_0^{2\pi} d\sigma'
e^{-{1\over 2}g(P+ig)\epsilon (\sigma'-\sigma )}
V^{(-1)}_{1}(\sigma')=e^{-{1\over 2}g(P+ig)}S(\sigma ),
$$
\hbox{where}\hfill
\beq
S(\sigma )=e^{2ih(\varpi +1)}\int_0^{\sigma } d\sigma' V^{(-1)}_{1}(\sigma')
+\int_{\sigma }^{2\pi } d\sigma' V^{(-1)}_{1}(\sigma')
\label{2.15}
\eeq
and similarly for the right-moving part. Notice, however, that here
everything is expressed in terms of the free field $\phi_1$ resp. $\bar
\phi_1$,
so that $\bar S(\sigma )$ differs from the corresponding expression in
ref.\cite{GSch} by the replacement $\bar \phi_2 \rightarrow \bar \phi_1 $.
The operators $V_{1/2}^{(1/2)}$ and $\bar V_{-1/2}^{(1/2)}$ formed from
the fields $\phi_2 ,\bar \phi_2$ are related to $V_{-1/2}^{(1/2)},
\bar V_{-1/2}^{(1/2)}$ by the application of $S$ resp. $\bar S$:
$$
V_{1/2}^{(1/2)}(\sigma )=N(\varpi ) V_{-1/2}^{(1/2)}(\sigma )S(\sigma )
$$
\beq
\bar V_{-1/2}^{(1/2)}(\sigma )=\bar N(\bar \varpi )
\bar V_{1/2}^{(1/2)}(\sigma )\bar S(\sigma )
\label{2.16}
\eeq
with (cf. Eq.\ref{A.7})\hfill
$$
N(\varpi )=\phantom{-}{1\over 2}{e^{-ih\varpi} \over \sin (h\varpi )}{\Gamma
[1+(1+\varpi )h/\pi ]\over \Gamma (1+h/\pi )
\Gamma (h\varpi / \pi)}
$$
\beq
\bar N(\bar \varpi )=-{1\over 2}{e^{-ih\bar \varpi} \over \sin (h\bar \varpi )}
{\Gamma
[1+(1-\bar \varpi )h/\pi ]\over \Gamma (1+h/ \pi)
\Gamma (-h\bar \varpi / \pi)}
\label{2.17}
\eeq
Hence, setting $\bar \varpi =\varpi $ again eventually,
the final form of $e^{-g\Phi}$ becomes
\beq
e^{\displaystyle -g\Phi (\tau ,\sigma)}= \sum_{m=\pm 1/2} a_m^{(1/2)}(\varpi )
|_{BCGT}
V_m^{(1/2)}(\tau +\sigma )\bar V_m^{(1/2)}(\tau -\sigma )
\label{2.18}
\eeq
with \hfill
$$
a_{+1/2}^{(1/2)}(\varpi )|_{BCGT} ={\zeta m\over 2}F(P+ig/2)e^{-gP/2}
{1\over N(\varpi )}
$$
\beq
a_{-1/2}^{(1/2)}(\varpi )|_{BCGT} ={\zeta m\over 2}F(P-ig/2)e^{-gP/2}
{1\over \bar N( \varpi )}
\label{2.19}
\eeq
We note that $V_{\pm 1/2}^{(1/2)}$ resp. $\bar V_{\pm 1/2}^{(1/2)}$
are nothing but the (normalized) quantum versions of the fields
$AA'^{-1/2}$ and $A$ resp. $B'^{-1/2}$ and $BB'^{-1/2}$, with $A(\tau +
\sigma )$ and $B(\tau -\sigma )$ the arbitrary functions parametrizing
the general classical solution
\beq
2g\Phi =\ln \left [{8A'B'\over \mu^2 (A-B)^2}\right ]
\label{2.19a}
\eeq
Let us now compare this with the corresponding expressions in the GN-G
construction. There we have for general $J$ \cite{G5}, ($\mu ^2 =1$)
$$
a_m^{(J)}(\varpi )|_{GN-G}= {(-1)^{J-m}\over \sqrt {\lfloor \varpi \rfloor
\lfloor \varpi +2m\rfloor }}\lambda_m^{(J)}(\varpi )N_m^{(J)}(\varpi )
\bar N_m^{(J)}(\varpi ),
$$
where \hfill
$$
\lambda_m^{(J)}(\varpi )= \left ( ^{2J}_{J+m} \right )
\prod_{r=0}^{2J} \lfloor \varpi +m-J+r\rfloor,
$$
$$
\left ( ^{2J}_{J+m} \right ) :={\prod_{r=1}^{J+m} \lfloor J-m+r\rfloor
\over \prod_{r=1}^{J+m} \lfloor r\rfloor },
$$
$$
N_m^{(J)}(\varpi )=\bar N_m^{(J)}(\varpi )={\prod_{r=1}^{2J} \Gamma (1+
rh/ \pi) \over \prod_{r=1}^{J-m} \Gamma (1+rh/ \pi)
\prod_{r=1}^{J+m} \Gamma (1+rh / \pi)} \times
$$
\beq
\prod_{r=1}^{J-m} \Gamma [(\varpi -r)h/ \pi]
\prod_{r=1}^{J+m} \Gamma [(-\varpi -r)h/ \pi]
\prod_{r=1}^{2m}{\sqrt{\Gamma [(\varpi +r-1)h/ \pi]}\over
\sqrt{\Gamma [(-\varpi -r)h/ \pi]}}
\label{2.20}
\eeq
For $J=1/2$, this reduces to
$$
a_{-1/2}^{(1/2)}(\varpi )|_{GN-G}=-\sqrt{\lfloor \varpi \rfloor \lfloor \varpi
-1
 \rfloor } \Gamma (-\varpi h/ \pi)\Gamma [(\varpi -1)h/ \pi]
$$
\beq
a_{+1/2}^{(1/2)}(\varpi )|_{GN-G}=+\sqrt{\lfloor \varpi \rfloor \lfloor \varpi
+1
 \rfloor } \Gamma (\varpi h/ \pi)\Gamma [(-\varpi -1)h/ \pi]
\label{2.21}
\eeq
The equivalence conditions Eq.\ref{1.4} read in this case
$$
a_{-1/2}^{(1/2)}(\varpi )|_{GN-G}={C_1 \over f_1(\varpi -1)}
a_{-1/2}^{(1/2)}(\varpi )|_{BCGT}
$$
\beq
a_{+1/2}^{(1/2)}(\varpi )|_{GN-G}=C_1 f_1(\varpi )a_{+1/2}^{(1/2)}(\varpi
)|_{BCGT}
\label{2.22}
\eeq
Eqs.\ref{2.22} are solved by
\beq
C_1=\pm {\sqrt 8 \Gamma (-h/\pi )\over \zeta },\quad f_1(\varpi )=\pm i
\label{2.23}
\eeq
The appearance of the imaginary unit is due to the fact that the hermiticity
properties of
$e^{{\textstyle -\alpha_-  \Phi_{GN}}}$ and $e^{{\textstyle -g
\Phi_{BCGT}}}$ are adjusted to the elliptic and
hyperbolic sector, respectively.
This completes the proof of equivalence of the BCGT and GN-G constructions
of  the inverse square root of the metric.
We mention that the same kind of equivalence
can be established for an earlier construction of $e^{-\alpha_-\Phi /2}$
in the GN-G formalism \cite{LuS}.

\section{ OW vs. GN-G}
Following ref.\cite{OW}, we start from the expression for the
Liouville exponentials given by Otto and Weigt\footnote
{In fact we consider here the expression given in the later papers of Weigt,
which differs from the original formula by a factor $(\sin (h)/h)^n$ in the
n-th term; the two versions are obviously equivalent in the sense of
Eq.\ref{1.4}. },
written
in a factorized form\footnote{In the first two papers of ref.\cite{OW},
the authors consider also slightly different normal ordering prescriptions
which do not allow a factorization into powers of screening charges. We
will not discuss these possibilities here.}
$$
e^{\displaystyle \lambda \varphi (\tau ,\sigma )}
=\sum_{n=0}^{\infty} {(-\mu ^2)^n  \lfloor 2\lambda\rfloor_n\over
\lfloor n\rfloor !}
Z_\eta^{(\lambda ,n)}
(\tau ,\sigma )
$$
with \hfill
$$
\lfloor 2\lambda\rfloor_n  \equiv \prod_{j=0}^{n-1} \lfloor 2\lambda +j\rfloor
\quad ,
\quad
\lfloor n\rfloor !\equiv
\prod_{j=1}^{n} \lfloor j\rfloor
$$
$$
Z_\eta^{(\lambda ,n)}(\tau ,\sigma )=F^{(\lambda ,n)}(\hat P+ih(n+\lambda))
\hat Z_\eta^{(\lambda ,n)}(\tau ,\sigma ),
$$
$$
F^{(\lambda ,n)}(\hat P+ih(n+\lambda))=
\prod_{j=0}^{n-1}
{1\over \sinh (\hat P -ih(j-n))\sinh (\hat P +ih(j+n+2\lambda ))}
$$
\beqa
\hat Z_\eta^{(\lambda ,n)}(\tau ,\sigma )=:e^{\lambda\eta\psi (\tau ,\sigma )}:
& \prod_{j=1}^n : \int_0^{2\pi}d\sigma_j d\bar \sigma_j
e^{\hat P(\epsilon (\sigma -\sigma_j)-\epsilon(\sigma -\bar \sigma_j ))}
\nnn
& e^{\eta\psi^+(\tau +\sigma_j )+\eta\psi^-(\tau -\bar \sigma_j )}:
\nnn
\label{3.1}
\eeqa
The normal ordering is the same as in the BCGT construction, and $h$ is the
same as in the GN-G notation. Table \ref{A.6} gives
for the oscillator part of
$e^{\eta\psi^+ +\eta\psi^-}$:
\beq
:e^{\eta\psi^+(\tau +\sigma_j )+\eta\psi^-(\tau -\bar \sigma_j )}:^{(osc)}=
\hat V_1^{(-1)}(\tau +\sigma_j ) \bar {\hat V}_1^{(-1)}(\tau -\bar \sigma_j )
\label{3.2}
\eeq
Notice the appearance of the operator $\bar V_1^{(-1)}$ in place of
$\bar V_{-1}^{(-1)}$ in the BCGT scheme, as a consequence of the use of the
free
field $\psi_{OW}=\ln A' +\ln{B'\over B^2}$ instead of $2g\psi_{BCGT} =\ln A'
-\ln B'$, with $A,B$ as in Eq.\ref{2.19a}.
The zero mode dependence of the $j$th factor can be written as
$$
:e^{\hat P (\epsilon (\sigma -\sigma_j)-\epsilon (\sigma -\bar \sigma_j ))}
e^{\eta\psi^+(\tau +\sigma_j)+\eta\psi^-(\tau -\bar\sigma_j )}:|_{\hbox{
zero mode}}=
$$
\beq
 e^{(\hat P +ih)(\epsilon (\sigma -\sigma_j )-\epsilon (\sigma -
\bar \sigma_j ))} e^{\hat Q +\hat P (\tau +\sigma_j)/\pi +\hat P
(\tau -\bar \sigma_j )/\pi}
\label{3.3}
\eeq
This is just the required dependence to complete $\hat V_1^{(-1)}$ and
$\bar {\hat V}_1^{(-1)} $ to $V_1^{(-1)}$ and $\bar V_1^{(-1)}$, if we again
take
into account the zero mode doubling as explained above. Thus we have
$$
\hat Z^{(\lambda ,n)}(\tau ,\sigma )=:e^{\lambda\eta\psi (\tau ,\sigma )}:
\prod_{j=1}^n\int_0^{2\pi}d\sigma_j d\bar \sigma_j e^{ih(\varpi +1)
(\epsilon (\sigma -\sigma_j )-\epsilon (\sigma -\bar \sigma_j ))}\times
$$
\beq
\phantom{\hat Z^{(\lambda ,n)}(\tau ,\sigma )=:e^{\lambda\eta\psi
(\tau ,\sigma )}:\prod_{j=1}^n\int_0^{2\pi}d\sigma_j d\bar \sigma_j}
V_1^{(-1)} (\tau +\sigma_j ) \bar V_1^{(-1)}(\tau -\bar \sigma_j )
\label{3.4}
\eeq
Introducing screening charges $S$, $\bar {\tilde S}$ with $S$ as in
Eq.\ref{2.15}
and
\beq
\bar {\tilde S} =e^{-2ih(\bar \varpi +1)}\int_0^\sigma d\sigma' \bar V_1^{(-1)}
(\sigma' ) +\int_\sigma^{2\pi}d\sigma' \bar V_1^{(-1)}(\sigma' ),
\label{3.5}
\eeq
we get
\beq
\hat Z^{(\lambda ,\eta )}(\tau ,\sigma )=:e^{\lambda\eta\psi (\tau ,\sigma )}:
(S\bar{\tilde S})^n
\label{3.6}
\eeq
It is equally easy to see that
\beq
:e^{\lambda \eta\psi (\tau ,\sigma )}: =V_\lambda ^{(-\lambda )}(\tau +\sigma )
\bar V_\lambda ^{(-\lambda )}(\tau -\sigma )
\label{3.7}
\eeq
Thus, with the identification $\lambda =-J$, the form of $e^{\lambda \varphi}$
agrees with the GN-G expression as written in the form of ref.\cite{GSch}.
As in ref.\cite{GSch}, we pass from $V_\lambda ^{(\lambda )} S^n$,
$\bar V_\lambda ^{(\lambda )}\bar{\tilde S}^n$ to the canonically
normalized operators $V_m^{(J)},\bar V_m^{(J)}$ with
\beq
\langle \varpi |V_m^{(J)} |\varpi +2m \rangle =1=
\langle \bar\varpi |\bar V_m^{(J)} |\bar\varpi +2m \rangle
\label{3.8}
\eeq
by means of the normalization factors $I_m^{(J)},\bar I_m^{(J)}$ given
in  appendix, so that
$$
e^{\displaystyle \lambda \varphi}
=\sum_{J+m=0,1,2,...}a_m^{(J)}(\varpi )|_{OW}
V_m^{(J)}\bar V_m^{(J)}$$
with\hfill
\beq
a_m^{(J)}(\varpi )|_{OW}={(-\mu^2)^n\lfloor 2\lambda\rfloor_n
\over \lfloor n\rfloor !}F^{(\lambda ,n)}(\hat P+ih(\lambda +n))
I_m^{(J)}(\varpi )\bar I_m^{(J)}(\varpi )
\label{3.9}
\eeq
Again, we have to find $C$ and $f(\varpi )$ relating the coefficients
$a_m^{(J)}$ of GN-G resp. OW. The calculation is straightforward and we
shall suppress the details. One finds
\beq
C_2=-{\Gamma (-h/\pi)\over 2} \qquad
f_2(\varpi )={1 \over 2}\sqrt{{\lfloor \varpi +1\rfloor\over
\lfloor \varpi \rfloor }}{\Gamma [1+(\varpi +1)h/\pi]\over
\Gamma (\varpi h/\pi)}
\label{3.10}
\eeq
The present discussion applies directly to the case where $J$ is
half-integer positive. However, it is in fact possible to generalize
to arbitrary real $J$. Indeed, it was shown in ref.\cite{GSch} that the GN-G
construction can be extended straightforwardly to continous $J$,
with  essentially the same expression to be used for the Liouville
exponentials as for the half-integer case (details will be given in a
forthcoming publication).
Correspondingly, Eq.\ref{3.10} possesses a natural continuation to arbitrary
$J$:
\beqa
\prod_{r=0}^{2m-1}\> f_2(\varpi +r)&=2^{-2m}\prod_{r=0}^{2m-1}\>\sqrt{
{\lfloor \varpi +r+1\rfloor \over \lfloor \varpi +r\rfloor}}
{\Gamma [1+(\varpi +r+1)h/\pi]\over \Gamma [(\varpi +r)h/\pi]}
\nnn
&=(h/ 2\pi)^{2m}\sqrt{{\lfloor \varpi +2m\rfloor\over
\lfloor \varpi \rfloor }}{\Gamma [(\varpi +2m)h/\pi]\over
\Gamma (\varpi h/\pi)}
\prod_{r=1}^{2m}\>(\varpi +r)\nnn
&=(h/ 2\pi)^{2m}
\sqrt{{\lfloor \varpi +2m\rfloor\over
\lfloor \varpi \rfloor }}{\Gamma [(\varpi +2m)h/\pi]\over
\Gamma (\varpi h/\pi)}{\Gamma (\varpi +2m+1)\over \Gamma (\varpi +1)}
\label{3.11}
\eeqa
The last expression makes sense for arbitrary $2m$, and we read off
\beq
S_2(\varpi )=\left ({2\pi\over h}\right )^\varpi {1\over
\sqrt{\lfloor \varpi\rfloor}
\Gamma (\varpi h/\pi)\Gamma (\varpi +1)}
\label{3.12}
\eeq
In ref.\cite{OW}, the locality of the Liouville exponentials was verified
only up to the third power of the cosmological constant; the above analysis
thus provides at the same time a proof to all orders that the result of
Otto and Weigt is correct.

A general word of caution should be added, however,
in the case of noninteger $2J$. The expansion Eq.\ref{1.2}
then becomes an infinite
sum which certainly does not converge in any naive sense. Consequently,
the equivalence of Eqs.\ref{1.1} and \ref{1.4} may be formal. Nevertheless,
it can be shown that whenever $a_m^{(J)}$ solves the locality conditions
\cite{GSch},
so will any other member of the equivalence class defined by transformations
of type Eq.\ref{1.4}. Thus, the GN-G resp. OW Liouville exponentials
are local as formal power series in the screening charges, as a consequence
of the underlying chiral algebra\cite{GN,G,GSch}.

\section{Group-theoretic approach to the
classical Liouville exponentials}
As a preparation of the coming section, it is pedagogical
to  temporarily return  to the classical case. Then we
have\footnote{In this section, we redefine the Liouville field, for
simplicity, so that the coupling constant need not be written any more.}
the general solution of ref.\cite{LS}:
$$
e^{\textstyle -j\Phi(z,\, \zb)}=
{<j,\, j| \Mb^{-1}(\zb) M(z) |j,\, j>\over (s(z) \sb(\zb))^j}
$$
\beq
{d M\over dz}=s(z) M j_-,\quad
{d \Mb\over d\zb}=\sb(\zb) \Mb j_+,
\label{4.1}
\eeq
where $s$ and $\sb$ are arbitrary functions of  a single variable.
The symbols  $j_\pm$ represent  $sl(2)$ generators satisfying
$[ j_+,\, j_-]=2 j_3$, and $|j,\, j>$ are highest-weight
states $j_+|j,j> =0$.  In Eq.\ref{4.1} and below, we use Euclidean
coordinates on the sphere,
$z=e^{\tau +i\sigma }$, $\zb =e^{\tau -i\sigma }$.
To establish the connection with the approaches of the last section,
note first that $s,\sb$ are related to the arbitrary functions
$A,B$
appearing in the general classical solution Eq.\ref{2.19a} by
\beq
s(z)=A'(z), \qquad \sb (\zb)=-B'(\zb)/B^2
\label{4.2}
\eeq
and thus $s$ and $\sb $ can be identified with the classical
equivalents of the screening charge densities $V_1^{(-1)}$, $\bar V_1^{(-1)}$
(up to normalization). It is then immediate to verify that Eq.\ref{4.1}
reduces to (the classical limit of) Eq.\ref{1.2} after evaluation of the
matrix element. Actually, the $SL(2,{\bf C})$ symmetry of the theory,
\beq
A \rightarrow {aA+b\over cA+d} \qquad B \rightarrow {aB+b\over cB+d},
\qquad ad-bc=1
\label{4.2a}
\eeq
allows us just as well to identify $s^{-1/2},\sb ^{-1/2}$ with any linear
combination of the quasiperiodic (Bloch wave) fields
$V_{-1/2}^{(1/2)}$ and $V_{+1/2}^{(1/2)}$ resp. $\bar V_{-1/2}^{(1/2)}$
and $\bar V_{+1/2}^{(1/2)}$. For the classical considerations below,
the assignment chosen is irrelevant; however, on the quantum level the
situation will be quite different.
Looking at Eq.\ref{4.1}, one may wonder
why  the Liouville exponential should be given by highest-weight
matrix elements. The basic reason is that
Eq.\ref{4.1} must be such that
\beq
e^{\textstyle -j_1\Phi(z,\, \zb)} e^{\textstyle -j_2\Phi(z,\, \zb)}
=e^{\textstyle -(j_1+j_2)\Phi(z,\, \zb)},
\label{4.3}
\eeq
by the very definition of the classical exponential function.
In order to verify this, we introduce the rescaled exponentials\footnote
{For the conceptual considerations here, the precise normalization of the
$E^{(j)}$ is not important. Therefore we don't specify the lower integration
limits
in Eq.\ref{4.4a}.}
$$
E^{(j)}(s,j_-)=s^{-j}(z)M(z)= s^{-j}(z)e^{\int^z s(z')dz' \ {\displaystyle
j_-}}
$$
resp.\hfill
\beq
\bar E^{(j)}(\sb ,j_+)=\sb ^{-j}(\zb)\bar M^{-1}(\zb)=
\sb ^{-j}(\zb)e^{-\int^\zb \sb (\zb')d\zb' \ {\displaystyle j_+}}
\label{4.4a}
\eeq
Thus the left-hand side of Eq.\ref{4.3} can be written as
$$
e^{\textstyle -j_1\Phi(z,\, \zb)} e^{\textstyle -j_2\Phi(z,\, \zb)}=
(<j_1,\, j_1| <j_2,\, j_2|)\bar E^{(j_1)}(\sb ,j_+) \otimes
\bar E^{(j_2)}(\sb ,j_+)\times
$$
$$
E^{(j_1)}(s,j_-) \otimes
E^{(j_2)}(s,j_-)(|j_1,\, j_1>|j_2,\, j_2>)
$$
$$
= (<j_1,\, j_1| <j_2,\, j_2|)\bar E^{(j_1+j_2)}(\sb ,j_+\otimes 1
+1 \otimes j_+)\times
$$
\beq
E^{(j_1+j_2)}(s,j_- \otimes 1 +1 \otimes j_-)
(|j_1,\, j_1>|j_2,\, j_2>)
\label{4.4}
\eeq
The highest weight-states considered are the only ones such that
the tensor product  gives a single  irreducible representation.
Its  spin is $j_1+j_2$, and  $|j_1,\, j_1>|j_2,\, j_2>$ is
the highest-weight vector. Since the matrix element of
$\bar M^{-1} M$ is  determined solely  by  the group structure, it only
depends upon the spin of the representation, and not upon the way it is
realized;
 hence Eq.\ref{4.3} follows.
In particular, we have
\beq
e^{\textstyle -j\Phi(z,\, \zb)}=
\left (e^{\textstyle -(1/2)\Phi(z,\, \zb)}\right)^{2j};
\label{4.5}
\eeq
and thus Eq.\ref{4.1} may be re-written using binomial coefficients
(more on this below).

It is well-known (see e.g. ref.\cite{GM})
that the Liouville solution and its Toda generalization
are actually tau-functions  in the sense of the Kyoto group\cite{JM}.
A characteristic feature of tau-functions is to involve
 highest-weight states. We shall not dwell into the precise
connection, since it is not directly evident from Eq.\ref{4.1}.
We shall rather recall  the existence of bilinear equations of the
Hirota type, which was the original motivation to introduce tau
functions.
The method of derivation we will use
 is not the same as the standard ones of
ref.\cite{JM} or ref.\cite{KW}. Its interest is that
 it will carry over to the
quantum case.

 One may obtain  a closed equation for the $j=1/2$
Liouville exponential as follows. Making use of
Eq.\ref{4.1}, let us compute  the antisymmetric
bilinear expression
$$
e^{ -\Phi/2} \partial_z \partial_\zb
e^{-\Phi/2} -
\partial_z e^{ -\Phi/2} \partial_\zb
e^{-\Phi/2}=
$$
$$
-<{1\over 2},\, {1\over 2} | \Mb^{-1} M |{1\over 2},\, {1\over 2}>
<{1\over 2},\, -{1\over 2} | \Mb^{-1} M |{1\over 2},\, -{1\over 2}>
$$
\beq
+<{1\over 2},\, -{1\over 2} | \Mb^{-1} M |{1\over 2},\, {1\over 2}>
<{1\over 2},\,  {1\over 2} | \Mb^{-1} M |{1\over 2},\, -{1\over 2}>
\label{4.5a}
\eeq
The point of this particular combination of
derivatives is that the functions $s$ and $\sb$ disappear, and
the result
is given by the matrix element of $\Mcb^{-1} \Mc$
in the $j=0$ representation, where
$\Mc=\Mcb=1$. Thus we get
\beq
e^{ -\Phi/2} \partial_z \partial_\zb
e^{-\Phi/2} -
\partial_z e^{ -\Phi/2} \partial_\zb
e^{-\Phi/2}=-1.
\label{4.6}
\eeq
Of course it is trivial to rederive this equation directly from the
Liouville equation; however
this form  -- the simplest example of
Hirota bilinear equations -- which only makes use of the
Liouville exponentials and not of the field $\Phi$ itself,  will
be much
easier to generalise to the quantum case. For later use we note
that this Hirota equation is equivalent to the following relation in the
Taylor expansion  for $z'\to z$, $\zb'\to \zb$.
\beq
e^{ -\Phi/2( z',  \zb')}
e^{ -\Phi/2(z ,\zb)} -
e^{ -\Phi/2( z',  \zb)}
e^{ -\Phi/2(z ,\zb')}\sim -(z'-z)(\zb'-\zb)
\label{4.7}
\eeq
Clearly we may obtain other bilinear equations for $\exp(-j \Phi)$
with $j\not=1/2$ by again projecting out the $j=0$ component of the
product.

\section{Liouville exponentials and q-deformations}

\subsection{The quantum group structure}
The deep connection of the quantum Liouville theory
with $U_q(sl(2))$ was there from the
beginning\cite{GN4}, but in disguise.
It was elucidated more recently in refs.\cite{B,G,CG,CGR1,CGR2}. The
quantum group-parameter $h$ is precisely the parameter\footnote{
There is actually a sort of doubling, and the complete structure is of
the type $U_q(sl(2))\odot U_\qhat(sl(2))$, corresponding to the
existence of more general operators $V_{m \mhat}^{J \Jhat}$. We do not
consider this possibility in this article. }
$2\pi \hbar \eta^2$  of GN-G.
Up to coupling constants, the fusing
and braiding matrices  of the $V_m^{(J)}$ fields are given by q-6j symbols.
However, the $V_m^{(J)}$ fields are not quantum group covariants; there
exists another basis of chiral vertex operators $\xi_M^{(J)}$,
related to the $V_m^{(J)}$
by a linear transformation, which transform as spin $J$ representations
of $U_q(sl(2))$.
Their fusing and braiding matrices are given by
q Clebsch-Gordan and universal R-matrix elements respectively.
In particular, the operator-product algebra of the $\xi_M^{(J)}$ corresponds to
making
q tensor products of representations.
We anticipate therefore that it is this basis which should be used when
trying to extend the considerations of the previous section to the
quantum level, so that the arbitrariness mentioned below Eq.\ref{4.2a} in the
precise assignment of $s(z),\sb (\zb)$ is lifted.
It was shown in ref.\cite{G5}
that  the Liouville
exponentials take the form ($2J$ positive integer)
\begin{equation}
e^{\textstyle -J\alpha_-\Phi(z, \zb  )}=
\sum _{M=-J}^J\> (-1)^{J+M}  \>e^{ih(J+M)}\>
\xi_M^{(J)}(z)\,
{\overline \xi_{-M}^{(J)}}(\zb)
\label{5.1}
\end{equation}
Since they play no role we do not indicate the factors $S(\varpi),S^{-1}
(\varpi )$
any more.  Of course they should be put back in order
to establish the precise
connection between this last formula and the expressions
of BCGT and OW.

The point of this section is to show that with  this last expression
for the Liouville exponential,  {\bf its quantum
properties are  directly connected with their
 classical analogues  by  standard q deformations}.

\subsection{Connection with q-binomials}

We shall use the notations of refs.\cite{G},\cite{CG}--\cite{CGR2}. One
introduces
\beq
\lfloor x \rfloor = {\sin (h x)\over \sin(h)}.
\label{5.2}
\eeq
The q-deformed binomial coefficients
 noted  ${P \choose Q}$ are defined by
\beq
{P \choose Q} := {\lfloor P \rfloor \! !
 \over \lfloor Q \rfloor \! !\lfloor P-Q \rfloor \! !},
\qquad \lfloor n \rfloor \! ! :=
\prod_{r=1}^n \lfloor r \rfloor,
\label{5.3}
\eeq
They are binomial coefficients for
 the expansion of $(x+y)^{2J}$, with
$x$ and $y$ non-commuting variables such that
\beq
xy=yx e^{-2ih}.
\label{5.4}
\eeq
 Indeed, it
is easy to verify that they satisfy
\beq
{m+1 \choose n}=e^{ihn}{m \choose n}+e^{-ih(m-n+1)}{m \choose n-1}.
\label{5.5}
\end{equation}
As a result one  sees that if one lets
\beq
(x+y)^N= \sum_{r=0}^N {N\choose r} e^{ ihr(N-r)} x^r y^{N-r}
\label{5.6}
\end{equation}
one has, as required,
\beq
(x+y)^{N+1}=(x+y) (x+y)^{N}.
\label{5.7}
\eeq

Recall the leading-order fusion of the $\xi$ fields\cite{G1}:
to leading order in the
short distance singularity at  $z' \to z$,
the product of $\xi$ fields behaves as\footnote{Recall we are working
here on the sphere, rather than on the cylinder.}
\beq
\xi_M^{(J)}(z)\,\xi_{M'}^{(J')}(z')
\sim
(z'-z)^{-2JJ'h/ \pi}
\>\lambda (J,M;\,J',M')
\,\xi_{M+M'}^{(J+J')}(\sigma ),
\label{5.8}
\eeq
\beq
 \lambda (J,M;\,J',M')=
\sqrt{ {  {2J \choose J+M}}\>
{ {2J' \choose J'+M'}}
\over {  {2J+2J' \choose J+J'+M+M'}} }
\>e^{ih(M'J-MJ')}.
\label{5.9}
\eeq
Thus, if we redefine
\beq
\eta_M^{(J)}\equiv \xi_M^{(J)}/ \sqrt{{2J \choose J+M}}
\label{5.10}
\eeq
we have
$$
\eta_M^{(J)}\,\eta_{M'}^{(J')}
\sim
\eta _{M+M'}^{(J+J')} \>e^{ih(M'J-MJ')}
$$
and thus \hfill
\beq
 \eta _M^{(J)}\sim (\eta _{1/2}^{(1/2)})^{J+M}(\eta _{-1/2}^{(1/2)})^{J-M}
e^{{ih\over 2}(J^2-M^2)}
\label{5.11}
\eeq
In the above formulae and hereafter the symbol $\sim$ means
leading term of the short-distance expansion, divided by the
singular short distance factor appearing in Eq.\ref{5.8}.
In terms of the
$\eta$ fields, the Liouville exponential takes the form

\beq
e^{\textstyle -J\alpha_-\Phi(z, \zb )}
=\qquad \sum _{J+M=0}^{2J} {2J \choose J+M} (-1)^{J+M} e^{ih(J+M)}\eta_M^{(J)}
(z)\etab _{-M}^{(J)}(\zb )
\label{5.12}
\eeq
or, using Eq.\ref{5.11},\hfill
$$
e^{\textstyle -J\alpha_-\Phi(z, \zb )} \sim \sum _{J+M=0}^{2J}
{2J \choose J+M} e^{ih(J+M)(J-M)}\times
$$
\beq
\left (-e^{ih}\eta_{1/2}^{(1/2)}\,
{\overline \eta_{-1/2}^{(1/2)}}\right )^{J+M}
\left (\eta_{-1/2}^{(1/2)}\,
{\overline \eta_{1/2}^{(1/2)}}\right )^{J-M}
\nnn
\label{5.13}
\eeq
which is completely analogous to the q-binomial expansion Eq.\ref{5.6},
if we identify $x=-e^{ih}\eta_{1/2}^{(1/2)}(z)\,{\overline \eta_{-1/2}^{(1/2)}}
(\zb )\, ,
\  y=\eta_{-1/2}^{(1/2)}(z)\, {\overline \eta_{1/2}^{(1/2)}}(\zb )$.
To avoid any possible confusion, we remark that Eq.\ref{5.4} does not imply
that the fields $\eta_{1/2}^{(1/2)}(z)\,{\overline \eta_{-1/2}^{(1/2)}}(\zb )$
and $\eta_{-1/2}^{(1/2)}(z')\, {\overline \eta_{1/2}^{(1/2)}}(\zb ' )$ commute
up to a factor; this is true only in the limit $z'\to z,\ \zb' \to \zb$.
 We remark
that Eq.\ref{5.12} can even be interpreted for arbitrary real $J$, in view
of the result of \cite{GSch} that the fields $V_m^{(J)}$ -- and hence
$\eta_m^{(J)}$ -- can be defined for any $J$. In this case, of course, the
sum in Eqs.\ref{5.12}, \ref{5.13} runs from zero to infinity.

\subsection{Liouville exponentials as quantum tau-functions}
As is well known,
the binomial coefficients  are closely related to representations of
$U_q(sl(2))$.  Consider group-theoretic states\footnote{We assume
for simplicity that $h/\pi$ is not rational.}
$\vert J, M >$, $-J\leq M\leq J$;
 together with operators $J_{\pm}$, $J_3$ such that:
\beq
J_\pm \vert J,M> =\sqrt{\lfloor J \mp M\rfloor
\lfloor J \pm M+1 \rfloor } \vert J, M\pm 1 >
\quad J_3 \vert J,M> =M\, \vert J,M>.
\label{5.15}
\eeq
These operators satisfy the $U_q(sl(2))$  commutation relations
\beq
\Bigl[J_+,J_-\Bigr]=\lfloor 2J_3 \rfloor, \quad
\Bigl[J_3,J_\pm \Bigr]=\pm J_\pm.
\label{5.16}
\eeq
It is elementary to derive the formulae
$$
<J,\, N\vert  (J_+)^P \vert J,M> =\sqrt{ \lfloor J+N\rfloor \! !
\lfloor J-M \rfloor \! ! \over \lfloor J-N\rfloor \! !
\lfloor J+M \rfloor \! !} \delta_{N,\, M+P}
$$
\beq
<J,\, N\vert  (J_-)^P \vert J,M> =\sqrt{ \lfloor J-N\rfloor \! !
\lfloor J+M \rfloor \! ! \over \lfloor J+N\rfloor \! !
\lfloor J-M \rfloor \! !} \delta_{N,\, M-P}
\label{5.17}
\eeq
Recall further that the co-products  of representations are  defined by
\beq
\Lambda (J)_\pm=J_\pm\otimes e^{ihJ_3}+
e^{-ihJ_3}\otimes J_\pm,\quad
\Lambda (J)_3=J_3\otimes 1+1\otimes J_3
\label{5.20}
\eeq
We will show now that the group-theoretical
classical formulae  of section 4 possess direct
quantum equivalents, obtained
by replacing $sl(2)$ by
$U_q(sl(2))$. We  start  from the general
classical solution and consider Eq.\ref{4.1} as the classical
tau function. The quantum tau function should then be given by a
representation of type Eq.\ref{4.1} of the operator $ e^{\textstyle
-J\alpha_-\Phi}$; that is, the q tau function should actually
be an operator instead of a function.
Let us introduce the (rescaled) q-exponentials
$$
E_q^{(J)}(\eta (z),J_-)=\sum_{J+M=0}^\infty \eta_M^{(J)}
{(J_-)^{J+M}\over \lfloor J+M\rfloor !}
$$
\beq
\bar E_q^{(J)}(\etab (\zb),J_+)=\sum_{J+M=0}^\infty e^{ih(J+M)}
(-1)^{J+M} \etab_{-M}^{(J)}
{(J_+)^{J+M}\over \lfloor J+M\rfloor !}
\label{5.22}
\eeq
which we take to be the quantum equivalents of the classical (rescaled)
exponentials $E^{(j)}(s,j_-)$, $\bar E^{(j)}(\sb ,j_+)$.
Indeed, in the
limit $h\to 0$, we see that $E_q^{(J)}(\eta ,J_-)$,
$\bar E_q^{(J)}(\etab ,J_+)$ reduce to $E^{(j)}(s,j_-)$,
$\bar E^{(j)}(\sb,j_+)$ if
we identify classically
$$
\eta_M^{(J)} =s^{-J}(z)(\int^z s(z')dz')^{J+M}=
\eta_{-J}^{(J)}(\eta_1^{(0)})^{J+M}=
\eta_{-J}^{(J)} \eta_{J+M}^{(0)}
$$
\beq
\etab_{-M}^{(J)} =\sb ^{-J}(\zb )(\int^\zb \sb (\zb')d\zb')^{J+M}
=\etab_{J}^{(J)}
(\etab_{-1}^{(0)})^{J+M}=
\etab_{J}^{(J)} \etab_{-J-M}^{(0)},
\eeq
which corresponds to the assignment
\beq
s(z)=\eta_1^{(-1)}(z) \qquad \sb (\zb )=\etab_{-1}^{(-1)}(\zb )
\eeq
Furthermore we will show now, following closely the calculation used
to derive
Eq.\ref{4.3} group-theoretically, that indeed
the $E_q^{(J)}$ obey a composition law appropriate for q-deformed
exponentials.
The argument was inspired by ref.\cite{GKL} (with an important difference-
see below).
In ref.\cite{CGR2}, the fusion algebra of the $\xi$ fields was determined
using the general scheme of Moore and Seiberg. One has
$$
\xi ^{(J_1)}_{M_1}(z_1)\,\xi^{(J_2)}_{M_2}(z_2) =
\sum _{J_{12}= \vert J_1 - J_2 \vert} ^{J_1+J_2}
g _{J_1J_2}^{J_{12}} (J_1,M_1;J_2,M_2\vert J_{12})\times
$$
\beq
\sum _{\{\nu\}} \xi ^{(J_{12},\{\nu\})} _{M_1+M_2}(z_2)
<\! <\!\varpi _{J_{12}},{\{\nu\}} \vert
V ^{(J_1)}_{J_2-J_{12}}
(z_1-z_2) \vert \varpi_{J_2}\! > \! >,
\label{5.29}
\eeq
where $\{\nu\}$ is a multi-index that labels
the  descendants, and $|\varpi _{J},{\{\nu\}} \! > \! >$ denotes
the corresponding state in the Virasoro Verma-module.
Similar equations hold for the $\xib$ fields.
The explicit expression of the  coupling constant $g$ is not needed
in the present argument. The symbol $(J_1,M_1;J_2,M_2\vert J_{12})$
denotes the q-Clebsch-Gordan coefficients.  It follows
from their  very definition (see, e.g. ref.\cite{G3})  that
$$
\sum_{M_1+M_2=M_{12}} (J_1,M_1;J_2,M_2\vert J_{12})
|J_1,\, M_1>\otimes |J_2,\, M_2>=
$$
\beq
\phantom{\sum_{M_1+M_2=M_{12}} (J_1,M_1;J_2,M_2\vert J_{12})
|J_1,\, M_1>}=
 {(\Lambda(J)_- )^{J_{12}-M_{12}}\over
\lfloor J_{12}-M_{12}\rfloor \! ! \sqrt{{2J_{12}\choose J_{12}-M_{12}}}}
 |J_{12},\, J_{12}>,
\label{5.30}
\eeq
and one finds
$$
E_q^{(J_1)}\! \left (\eta(z_1),\,
J_-\otimes 1
 \right )
E_q^{(J_2)}\! \left (\eta(z_2),\,
1\otimes  J_-\right )
(\vert J_1,\, J_1 >\vert J_2,\, J_2 >)=
$$
$$
\sum _{J_{12}= \vert J_1 - J_2 \vert} ^{J_1+J_2}
g _{J_1J_2}^{J_{12}} \times
$$
\beq
\sum _{\{\nu\}} E_q^{(J_{12} ,\{\nu\})}\! \left (\eta(z_2),\,
\Lambda(J)_- \right )|J_{12},\, J_{12}>
<\! <\!\varpi _{J_{12}},{\{\nu\}} \vert
V ^{(J_1)}_{J_2-J_{12}}
(z_1-z_2) \vert \varpi_{J_2}\! > \! >
\label{5.31}
\eeq
In particular, to leading order one has
$$
E_q^{(J_1)}\left (\eta(z_1),\,
J_-\otimes 1
 \right )
E_q^{( J_2)}\left (\eta(z_2),\,
1\otimes  J_-\right )
(\vert J_1,\, J_1 >\vert J_2,\, J_2 >)\sim
$$
\beq
E_q^{(J_{1}+J_2 )}\left (\eta(z_2),\,
\Lambda(J)_- \right )|J_{1}+J_2,\, J_{1}+J_2>.
\label{5.32}
\eeq
which is the natural multiplication law for
q-exponentials involving quantum group generators.
The coproduct is non-symmetric between the two
representations. On the left hand side this
comes from the non-commutativity   of the $\eta $ fields as quantum
field operators. Thus the present definition  of q exponentials
is conceptually
different from the usual one where the group ``parameters''
are c numbers. Let us recall the latter for completeness. If on defines
\beq
e_q(X)\equiv \sum_{r=0}^\infty
 {X^r\over \lfloor r\rfloor\! !} e^{-ihr(r+1)/2},
\label{x.17}
\eeq
one has
\beq
e_q(x J_\pm \otimes e^{ihJ_3})
e_q(x e^{-ihJ_3} \otimes J_\pm )= e_q(x \Lambda (J)_\pm).
\label{x.20}
\eeq
Since they transform the q sum of infinitesimal generators into
products, the q exponentials are the natural  way to exponentiate
q Lie algebras. The last equation should be compared with Eq.\ref{5.32}.
Now the non-symmetry of the co-product is taken care of by
exponentiating non-commuting group elements ($J_\pm \otimes e^{ihJ_3}$,
and $e^{-ihJ_3} \otimes J_\pm$ on the left-hand side), with x a number.
This is in contrast with Eq.\ref{5.32}, where $J_\pm \otimes 1$, and
$1 \otimes J_\pm$ are used instead.

Finally, we can rewrite Eq.\ref{5.12} under the
form\footnote{Related formulae are already given in ref.\cite{LS2},
p. 27.}
\begin{equation}
e^{\textstyle -J\alpha_-\Phi(z, \zb )}=
<J,\, J \vert \bar E_q^{(J)}(\etab(\zb ),\,   J_+)
E_q^{(J)}(\eta(z),\, J_-) \vert J,\, J>=\tau_q (\eta ,\bar\eta ).
\label{5.25}
\end{equation}
Indeed we will see that it
 possesses the obvious q-analogues of the properties Eq.\ref{4.4}
and Eq.\ref{4.5}. Hence,
Eq.\ref{5.25} should be viewed as the quantum version of the
Leznov-Saveliev
formula Eq.\ref{4.1}.
Since $\eta$ and $\etab$ commute, we deduce that
$$
e^{\textstyle -J_1\alpha_-\Phi(z_1, \zb_1 )}
e^{\textstyle -J_2\alpha_-\Phi(z_2, \zb_2 )}=
$$
$$
(<J_1,\, J_1 \vert <J_2,\, J_2 \vert)
 \bar E_q^{(J_1)}\! \left (\etab(\zb_1),\,
J_+\otimes 1\right)
\bar E_q^{(J_2)}\! \left (\etab(\zb_2),\,
1\otimes  J_+\right  )
$$
\beq
E_q^{(J_1)}\! \left (\eta(z_1),\,
J_-\otimes 1
 \right )
E_q^{(J_2)}\! \left (\eta(z_2),\,
1\otimes  J_-\right )
(\vert J_1,\, J_1 >\vert J_2,\, J_2 >).
\label{5.28}
\eeq

Making use of the fusion algebra Eq.\ref{5.29}, together with its
counterpart for the $\bar \eta$ fields gives back the fusion algebra
of the Liouville exponentials derived in ref.\cite{G5}. To leading
order one has
\beq
e^{\textstyle -J_1\alpha_-\Phi}
e^{\textstyle -J_2\alpha_-\Phi}\sim
e^{\textstyle -(J_1+J_2)\alpha_-\Phi}.
\label{zzz}
\eeq
Clearly these properties are natural generalizations of the
classical features recalled in section 4.

\subsection{The quantum Hirota equation}

We retain the basic idea of the calculation which led to
Eq.\ref{4.6} or \ref{4.7}.
Clearly, Eq.\ref{4.7}  should  be replaced by a relation
for the operator-product expansion.
  The OPE of the
$J=1/2$ Liouville
exponential  has the form\cite{G5}
$$
e^{\textstyle -{1\over 2}\alpha_-\Phi(z_1, \zb_1 )}
e^{\textstyle -{1\over 2}\alpha_-\Phi(z_2, \zb_2 )} \sim
\left[ (z_1-z_2)(\zb_1-\zb_2)\right ] ^{-h/2\pi}
 e^{\textstyle -\alpha_-\Phi(z_2, \zb_2 )}
$$
\beq
+\left[(z_1-z_2)(\zb_1 -\zb_2) \right ] ^{1+3h/2\pi}
c_0+\> \hbox{descendants},
\label{5.33}
\eeq
where $c_0$ is a constant that can be changed by a global shift
of the Liouville field.
The second Liouville exponential is equal to a constant, since its spin
$J$ is equal to zero. This property is the quantum equivalent of
Eq.\ref{4.7}. Thus, the quantum Hirota equation simply
results  from the fact that
for $J=0$, $\exp( -J\alpha_-\Phi)=\hbox{ cst}$.
One sees that choosing a particular combination of derivatives
in a bilinear classical expression
of $J=1/2$
Liouville exponentials  is replaced by  picking up the spin zero
term in the operator-product expansion of $\exp( -(1/2)\alpha_-\Phi)$
with itself.
Note that due to the
quantum effects, the difference of the powers of
$(z_1-z_2)(\zb_1 -\zb_2)$
between the first and second term is not equal to one --- it is
equal to $1+2h$ ---   so that a simple antisymmetrization is not enough,
as in the classical case,  to remove the first term.
Clearly, Eq.\ref{5.33} gives non-trivial equations relating the
matrix elements of the quantum Liouville exponentials.
\section{Conclusions}
Starting from three manifestly different frameworks for the quantum
Liouville theory, we have shown that, apart from mere notational
differences, the Liouville exponentials obtained from them coincide up
to a simple equivalence transformation. This implies in particular
that in all cases the quantum structure is determined by the underlying
$U_q(sl(2))$ symmetry, though this is manifest only in the GN-G approach.
Note  that  in the work of GN-G, the $U_q(sl(2))$ symmetry
was not used as an input to define the quantum construction, but arose
as a consequence of the fundamental requirements of locality and conformal
invariance. Together with the results obtained here this provides a
very strong
indication that the Liouville quantum theory is unique.

The frameworks of BCGT and OW were formulated only
for the hyperbolic sector
corresponding to regular solutions of the Liouville equation and real
zero modes. However, their equivalence with the GN-G theory in this sector
shows how they should be extended to the elliptic sector with imaginary
zero modes, since the GN-G operators by construction apply simultaneously
to both cases. As regards correlation functions, an equivalence of type
Eq.\ref{1.1} implies, in principle,
  that the corresponding n-point functions
should agree, possibly up to an overall n-independent factor related to the
treatment of the end points\cite{G5}. At present, however,
this  is completely clear
only in the case $J$ half-integer positive where the Liouville
exponentials can be represented as {\it finite} sums over chiral primaries.
In the general case of arbitrary $J$, the meaning of the infinite sum has
yet to be clarified and the equivalence of Eqs.\ref{1.1} and \ref{1.4}
is nontrivial; this is the subject of ongoing investigations.

If the quantum Liouville theory is indeed unique, and  governed by the
$U_q(sl(2))$ quantum group symmetry, then it is natural to expect that
 it should be given
by a q-deformation of the group-theoretic approach of Leznov and Saveliev
to the classical theory. We have shown that in fact the quantum version
of the Leznov-Saveliev formula can be obtained by a combination of standard
q-deformation techniques and results of the quantum group analysis of
\cite{G}. It turns out that the fusion properties of the Liouville
exponentials and their chiral constituents can be neatly described in terms
of the multiplication law for field theoretic
q-exponentials of a new type. We also presented an
interpretation of the quantum Leznov-Saveliev representation in terms
of q tau functions which satisfy a bilinear Hirota-type equation,
for the simplest case $J=1/2$.
Defining the quantum equations of motion in this way,
rather than by a direct quantization of the classical Liouville
equation as in \cite{BCGT,OW}  has the obvious advantage that only
conformal objects are involved. On the other hand,
one expects in general  (see ref.\cite{GKL}, for instance)
that the q-Hirota equations should take the form of difference equations,
which, however, is not immediately evident from Eq.\ref{5.33}.
 It would be desirable to have a better
understanding of this point, and also to generalize the above picture
to the case of arbitrary Toda theories.
\vskip 5mm
\noindent
{\bf Acknowledgements}

The authors are indebted to Dimitri Lebedev for his detailed description
of his unpublished results,   and to   Gerhard Weigt for useful
explanations about  his work. This work was supported, in part by
the EEC grant \# SC1*-0394-C.
 \begin{appendix}

\section{ appendix}
{\bf {\large Notations}}
$$
\begin{array}{rrr}
\hbox{\large GN-G } \qquad\qquad &\qquad \qquad\hbox{\large BCGT}\qquad\qquad
 &\qquad\qquad \hbox{\large OW}
\end{array}
$$
\beqa
c=1+{6\pi\over h}(1+{h\over\pi})^2 \qquad\qquad &c=1+{2\pi\over g^2}
(1+{g^2\over 2\pi})^2\qquad\qquad
& c=1+{48\pi \over \gamma^2}\nnn
h\equiv 2\pi \hbar\eta^2, \hbar ={3\over c-1} \qquad\qquad
&g=\sqrt{2h} \qquad\qquad &\gamma =
{2\sqrt{2h}\over 1+{h\over \pi}} \nnn
\eta =\eta_- ={1\over 4\hbar}(1-\sqrt{1-8\hbar})\qquad\qquad
&\zeta ={\sqrt{2\sin g^2/2}\over g},\  8m^2=\mu^2 \qquad\qquad &
\phantom{\gamma }=4\sqrt{\pi\hbar}\nnn
\qquad\qquad &(\mu^2 =\hbox{cosm. constant})\nnn
\label{A.1}
\eeqa
\noindent
{\bf{\large Liouville exponentials}}\hfill\break
\vskip 0.2cm
\noindent
\qquad\qquad\quad{\large GN-G }\footnote{We use here the notation
of the more recent papers of Gervais \cite{G}. The OW notation in fact
agrees directly with the older GN notation}
 \qquad\qquad \qquad \qquad{\large BCGT}\qquad\qquad
 \quad\qquad\qquad {\large OW}

\beqa
e^{ {\displaystyle -J\alpha_-\Phi}}\qquad\qquad
& e^{ {\displaystyle \beta \Phi}}\qquad \qquad&
e^{ {\displaystyle \lambda  \varphi }}\nnn
(\alpha_- =\sqrt{{2h\over \pi}})\qquad\qquad\nnn
\Delta =-J-{h\over \pi}J(J+1)\qquad\qquad &\Delta ={\beta\over 2g}
(1+{g^2\over 2\pi})-{\beta^2\over 8\pi}\qquad\qquad &\Delta =
-{1\over 2}(\lambda \eta +\hbar \lambda^2\eta ^2)\nnn
\qquad\qquad &\Phi_{BCGT}={ \alpha_- \over 2g}\Phi_{G}\qquad\qquad &\varphi
=\alpha_-\Phi_{G}\nnn
\qquad\qquad &\beta=-2gJ \qquad\qquad &\lambda =-J \nnn
\sqrt{g}=e^{{\displaystyle \alpha_-\Phi}} \qquad\qquad
&\sqrt{g}=e^{{\displaystyle 2g\Phi}} \qquad\qquad
&\sqrt{g}=e^{{\displaystyle \varphi}} \nnn
\label{A.2}
\eeqa

\noindent
{\bf{\large  Free fields}}\hfill\break
\vskip 0.2cm
{\large GN-G  :}\hfill \break
$$
\phi_j(u)=q_0^{(j)}+p_0^{(j)}u +i\sum_{n\ne 0}e^{-inu}{p_n^{(j)}\over n}
$$
$$
\bar\phi_j(v)=\bar q_0^{(j)}+\bar p_0^{(j)}v +i\sum_{n\ne 0}e^{-inv}
{\bar p_n^{(j)}\over n}
$$
with\hfill
$$
u=\tau +\sigma ,\quad v=\tau -\sigma ,\quad j=1,2 \quad ,
$$
$$
[q_0^{(j)},p_0^{(j)}]=i=[\bar q_0^{(j)},\bar p_0^{(j)}], \quad
[p_n^{(j)},p_m^{(j)}]=n\delta_{n,-m}=[\bar p_n^{(j)},\bar p_m^{(j)}]
$$
It is often convenient to use the rescaled zero modes \hfill
$$
\varpi := {i\over \sqrt \hbar \eta}p_0^{(1)} \qquad \bar \varpi :=
{i\over \sqrt \hbar \eta}\bar p_0^{(1)}
$$
The above fields are related to the functions $A(u),B(v)$ describing the
general
solution of the Liouville equation by
\footnote{We remind the reader of the remarks made below Eq.\ref{2.9}
about the doubling of the zero modes in the GN-G  notation}
$$
\phi_1={1\over \sqrt \hbar}\ln A'^{-1/2} \qquad
\phi_2={1\over \sqrt \hbar}\ln AA'^{-1/2}
$$
$$
\bar \phi_1 = -{1\over \sqrt \hbar }\ln B'^{-1/2} \qquad
\bar \phi_2 =-{1\over \sqrt \hbar} \ln BB'^{-1/2}
$$
\beq
\label{A.3}
\eeq
up to additive constants depending only on the zero modes $p_0,\bar p_0$.
\vskip 1.5cm
\noindent
{\large BCGT :}\hfill\break
$$
\psi (\tau ,\sigma )=Q+P{\tau \over 2\pi}+\psi_L(u) +\psi_R(v) \quad ,
\quad \tilde \psi (\tau ,\sigma )=\psi_L(u)-\psi_R(v)
$$
with\hfill
$$
\psi_L(u)={i\over \sqrt{4\pi}}\sum_{n\ne 0}{A_n \over n}e^{-inu} \quad
\psi_R(v)={i\over \sqrt{4\pi}}\sum_{n\ne 0}{B_n \over n}e^{-inv}
$$
$$
[Q,P]=i \quad [A_n,A_m]=n\delta_{n,-m}=[B_n,B_m]
$$
relation to GN-G :\hfill
$$
\psi_L(u)=-{1\over\sqrt{4\pi}}\phi_1^{(osc)}(u) \quad
\psi_R(v)=-{1\over\sqrt{4\pi}}\bar\phi_1^{(osc)}(v)
$$
$$
P=-\sqrt{4\pi}p_0^{(1)} \quad Q=-{1\over \sqrt{4\pi}}q_0^{(1)} \quad A_n=
-p_n^{(1)} \quad B_n=-\bar p_n^{(1)} \ \, (n\ne 0)
$$
($\phi_1^{(osc)} \equiv$ oscillator part of $\phi_1$).\hfill\break
$\psi (\tau ,\sigma )$ is related to the functions $A(u)$ and $B(v)$ by
$$
2g\psi (\tau ,\sigma ) =\ln A'(u) -\ln B'(v)
$$
\beq
\label{A.4}
\eeq
\vskip 1.0cm
\noindent
{\large OW:}\hfill\break
$$
\psi (\tau ,\sigma )=\psi^+(u)+\psi^-(v) \quad ,
$$
with\hfill
$$
\psi^+(u)=\gamma ({1\over 2}Q+{1\over 4\pi}Pu +{i\over \sqrt{4\pi }}
\sum_{n\ne 0}{a_n^+ \over n}e^{-inu})
$$
$$
\psi^-(v)=\gamma ({1\over 2}Q+{1\over 4\pi}Pv +{i\over \sqrt{4\pi }}
\sum_{n\ne 0}{a_n^- \over n}e^{-inv})
$$

$$
[Q,P]=i \quad [a_n^+,a_m^+]=n\delta_{n,-m} =[a_n^-,a_m^-]
$$
relation to GN-G :\hfill
$$
\psi^{+(osc)}(u)=-2\sqrt{\hbar} \phi_1^{(osc)}(u) \quad
\psi^{-(osc)}(u)=+2\sqrt{\hbar} \bar\phi_2^{(osc)}(v)
$$
$$
P=-\sqrt{4\pi}p_0^{(1)} \quad Q=-{1\over \sqrt{4\pi}}q_0^{(1)} \quad
a_n^{+}=-p_n^{(1)} \quad a_n^{-}=+\bar p_n^{(2)} \ \, (n\ne 0)
$$
It is convenient to use the abbreviations \cite{OW}
$$
\hat P\equiv \sqrt{\pi\hbar} \eta P=-2\pi\sqrt{\hbar}\eta p_0^{(1)},
\quad \hat Q\equiv  4\sqrt{\pi\hbar}\eta Q=-2\sqrt{\hbar}\eta q_0^{(1)},
$$
$$ [\hat Q,\hat P]=2ih$$
$\psi (\tau ,\sigma )$ is related to the functions $A(u)$ and $B(v)$ by \hfill
$$
\psi (\tau ,\sigma ) =\ln A'(u) +\ln {B'(v) \over B^2}
$$
\beq
\label{A.5}
\eeq

\noindent
{\bf{\large Free field exponentials}}\hfill\break
$$
\begin{array}{rrr}
\hbox{\large GN-G}   \qquad\qquad &\qquad \qquad\hbox{\large BCGT}\qquad\qquad
 &\qquad\qquad \hbox{\large OW}
\end{array}
$$
\beqa
V_{-1/2}^{(1/2)}(u)=e^{+\sqrt{h/ 2\pi}{\displaystyle \phi_1(u)}}
\qquad\qquad\qquad
& e^{{\displaystyle -g\psi_L}} \qquad\qquad \qquad\qquad\
& e^{ {\displaystyle -\eta \psi^+ /2}}\qquad\quad\nnn
V_{+1/2}^{(1/2)}(u)=e^{-\sqrt{h/ 2\pi}{\displaystyle \phi_2(u)}}
\qquad\qquad\qquad
& {\displaystyle  /}\qquad\qquad\qquad\qquad\  &{\displaystyle  /} \nnn
\bar V_{-1/2}^{(1/2)}(v)=e^{-\sqrt{h/ 2\pi}\bar {\displaystyle \phi_2(v)}}
\qquad\qquad
\qquad
& {\displaystyle  /}\qquad\qquad\qquad\qquad \  &e^{
{\displaystyle -\eta  \psi^- /2}}\nnn
\bar V_{+1/2}^{(1/2)}(v)=e^{-\sqrt{h/ 2\pi}{\displaystyle \bar \phi_1(v)}}
\qquad\qquad
\qquad
&e^{{\displaystyle g\psi_R}} \qquad\qquad \qquad\qquad\ &{\displaystyle  /}\nnn
V_{+1}^{(-1)}(u)=e^{-2\sqrt{h/ 2\pi}{\displaystyle \phi_1(u)}}
\qquad\qquad\qquad
&e^{{\displaystyle 2g\psi_L}} \qquad\qquad \qquad\qquad\ &e^{
{\displaystyle \eta \psi^+}}\nnn
V_{-1}^{(-1)}(u)=e^{-2\sqrt{h/ 2\pi}{\displaystyle \phi_2(u)}}
\qquad\qquad\qquad
& {\displaystyle  /} \qquad\qquad\qquad\qquad\  &{\displaystyle  /}\nnn
\bar V_{+1}^{(-1)}(v)=e^{+2\sqrt{h/ 2\pi}{\displaystyle \bar\phi_2(v)}}
 \qquad\qquad\qquad
&{\displaystyle  /} \qquad\qquad\qquad\qquad\  &e^{ {\displaystyle \eta
\psi^-}}\nnn
\bar V_{-1}^{(-1)}(v)=e^{+2\sqrt{h/ 2\pi}{\displaystyle \bar\phi_1(v)}}
 \qquad\qquad\qquad
&e^{{\displaystyle -2g\psi_R}} \qquad\qquad\qquad\qquad\  &{\displaystyle
/}\nnn
\label{A.6}
\eeqa
The above correspondences are identities for the oscillator parts of the
operators; the zero mode dependences are discussed in the text.
Furthermore, we remark that we can replace everywhere $\phi_1$ by $\phi_2$
since this is nothing but a particular SL(2) transformation (corresponding
to the exchange $A\rightarrow -1/A,\ B\rightarrow -1/B$ in the classical
solution). The particular assignment we have chosen above in comparing
the free fields resp. free field exponentials is thus to this extent a
matter of convention.

Finally, we note that the relation between the
fields $V_\lambda ^{(-\lambda )}S^n $ and the normalized operators
$V_m^{(J)},\ J=-\lambda ,\, m=n-J$ is given by
$$
V_\lambda ^{(-\lambda )}S^n =I_m^{(J)}(\varpi )V_m^{(J)}
$$
with \hfill
$$
I_m^{(J)}(\varpi )=i^n \prod_{l=1}^n \left \{e^{i\pi  \beta (l-1)}
(1-e^{2\pi i (\gamma +  \beta (l-1))})\right \} \times
$$
$$
\prod_{l=1}^n \left \{{\Gamma (1-\beta ) \Gamma (1+\gamma +(l-1)\beta )
\Gamma (1+\alpha +(l-1)\beta )
\over \Gamma (1-l\beta ) \Gamma (2+\gamma +\alpha +(n-2+l)\beta )}\right \}
$$
\beq
\alpha = 2J{h \over \pi}, \quad
\beta = -{h \over \pi}, \quad
\gamma={h \over \pi} (\varpi +2m -1) -1.
\label{A.7}
\eeq
A completely analogous relation connects $\bar V_\lambda ^{(-\lambda )}
\bar{\tilde S}^n $ with $\bar V_m^{(J)}$, the only change being that
$\varpi $ is to be replaced by $\bar \varpi $ and the imaginary unit $i$
by $-i$.
\end{appendix}

\end{document}